\documentclass[11pt,twocolumn,english,review]{IEEEtran}
\usepackage[latin9]{inputenc}
\usepackage[a4paper]{geometry}
\usepackage{array}
\usepackage{float}
\usepackage{graphicx}
\usepackage[numbers]{natbib}

\makeatletter

\providecommand{\tabularnewline}{\\}

\usepackage{gensymb}
\usepackage{circuitikz}

\usepackage{lineno}

\makeatother

\usepackage{babel}
\begin{document}
\author{     
	\IEEEauthorblockN{G. Orr\IEEEauthorrefmark{1}, G. Golan\IEEEauthorrefmark{2}\thanks{\IEEEauthorrefmark{1}Corresponding author, gilad.orr@ariel.ac.il}}
    \\\IEEEauthorblockA{\IEEEauthorrefmark{1}Department of Physics, Ariel University, Ariel 40700, Israel}     
	\\\IEEEauthorblockA{\IEEEauthorrefmark{2}Department of Electrical Engineering, Ariel University, Ariel 40700, Israel}
}
\title{Crystalline quality in aluminium single crystals, characterized by
X-Ray diffraction and Rocking-Curve analysis}
\maketitle
\begin{abstract}
Aluminum single crystals are tested using X-Ray Bragg diffraction,
which may have applications in microscopy and electronics fabrication
industry. Yet, their efficiency for x-ray beam diffraction depends
on the accurate crystal orientation, the microstructure and imperfections.
Moreover, the final sample that is formed from the as-grown crystal
by cutting, grinding, polishing and chemical etching, introduces various
surface defects that penetrate deep into the crystal affecting its
natural structure. Defect penetration is attributed to the fact that
ultra-pure aluminum single crystals are soft and ductile with a hardness
in the range of 2\textasciitilde 2.5 mho. This leads to lattice deformation,
resulting in a deviation from the crystallographic orientation of
the final device, affecting the diffraction intensity and an apparent
shift in the Bragg angle. In this work we investigate the influence
of processing aluminum single crystals by mechanical and chemical
means using XRD and Rocking-Curve broadening as a quantitative indication
concerning the depth of the damage. This is a preliminary step in
supporting future work on the study of electrical conduction in aluminum
single crystals. Supplementing electrical conductivity measurements
of aluminum, quality assessment of defects in front cell aluminum
conductors can assist in designing novel low resistance aluminum conductors
replacing the currently widely used and relatively rare silver.
\end{abstract}


\maketitle

\section{Introduction}

Conduction in elemental metallic crystals is still intensively investigated
\citep{lapovok2020enhancement} and is of great technological importance
in modern microelectronics and photovoltaics. Of the FCC metallic
crystal family, aluminum is widely used in modern microelectronic
fabrication as it is a relatively abundant, acceptable conductor which
to some extent is superior to copper for it does not diffused into
silicon at operating temperatures $(<450\degree C)$. That said, with
all the complications copper adds to device manufacturing aluminum's
conductivity is only 63\% compared to the conductivity of copper ($0.0172\,\Omega\cdot mm^{2}/m$
vs. $0.0282\,\Omega\cdot mm^{2}/m$). Thus increasing aluminum's conductivity
has its merits. Current photovoltaic cell bus bar and finger technology
consists of silver screen printing thus relying on a relatively rare
and expensive metal. Replacing the silver with an abundant metal may
reduce the production costs considerably, therefore given the considerations
above aluminum is a good candidate. Depositing the aluminum on the
silicon substrate does not require a conductive seed layer or a Ni
barrier layer (as required for Cu), as it not a deep level impurity,
further simplifying the process and reducing its cost. One veteran
method which has re-emerged is metal plating, with recent work demonstrating
its viability \citep{ricci2021light}. Checking on Dimensions \citep{Dimensions}
shows that since the time of peak interest during 2012 there are on
average 600 annual publications concerning using aluminum for front
metallization of solar cells. As we have seen Al is inferior to Ag
and Cu concerning its resistivity adding to a significant power efficiency
loss due to series resistance resulting from the bus bars and fingers
on the front side of the cell. Therefore further study is required
in order to reduce the resistance of the Al front conductors. Lapovok
et al. \citep{lapovok2020enhancement} demonstrated a 3.7\% improvement
(reduction) in resistivity from $\rho=2.96\,\mu\Omega\cdot cm$ to
$\rho=2.85\,\mu\Omega\cdot cm$ following annealing at $600\degree C$
for 48h. Following a similar treatment in a boron rich environment
resulted in the resistivity dropping to $\rho=2.62\,\mu\Omega\cdot cm$
i.e. a 11.5\% in resistivity which is a significant improvement. This
increase was attributed to reduced scattering centers due to a gettering
effect and reduced defects due to the additional annealing. Increasing
the layers height resulting in an increase of the conductors cross
section may assist in reducing the series resistance though the geometry
may need some redesigning to avoid shadowing. Thus understanding mechanisms
of electrical conduction in single crystals can lead to methods for
reducing the electrical resistance of the front solar cell conductors.
Ultra-high-purity aluminum single crystals are a good starting point
for such experimental work, and high resolution X-ray diffraction,
a measurement technique for assessing crystal quality and defects.
However, it is yet a significant challenge to determine the degree
of crystalline perfection that will provide the optimal conditions
for a significant improvement compared to the current accepted values.
Every stage of the fabrication process has to be considered for its
contribution to defect formation and elimination. Growing Aluminum
single crystals was mostly investigated during several decades at
the mid of the 20th century \citep{mohanlal1982crystal} by the well-known
Bridgman method \citep{zerbib2020open} or by using the grain growth
technique. 
\begin{figure}[h]
\begin{centering}
\includegraphics[angle=-90,scale=0.065]{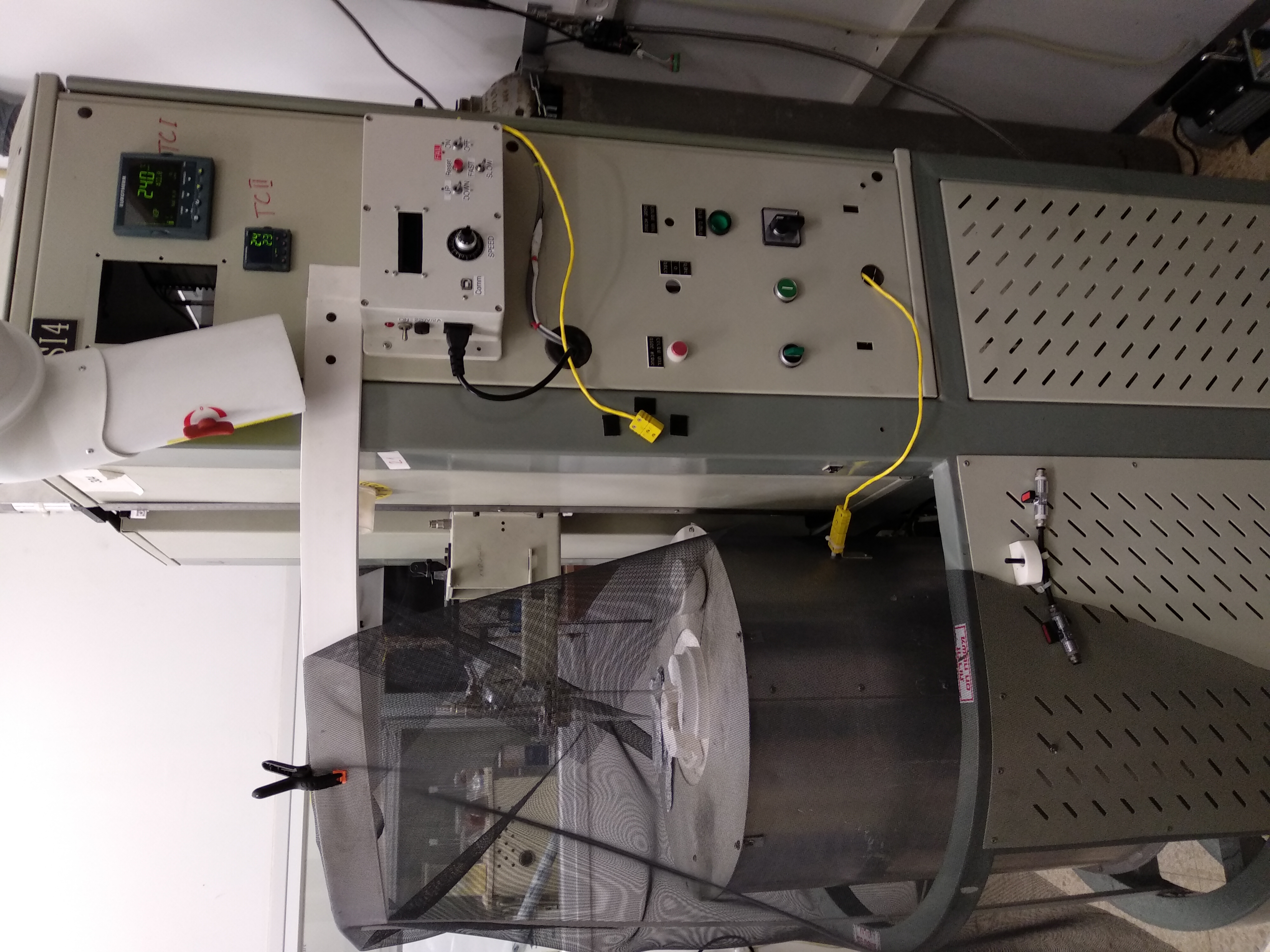}
\par\end{centering}
\caption{\label{fig:Bridgman-based-growth}Bridgman based growth furnace with
an innert growth environment}

\end{figure}
Recently, some new methods have been applied for growing single crystal
foils of several metals \citep{jin2018colossal} that can be used
at an industrial scale. Yet, the deformation in the crystals was mostly
studied by intentional stressing of the crystal in order to perform
the deformation, as well as studying the effects of heat treatment
and annealing processes and characterize the resulting microstructure
by means of X-ray diffraction methods and high-resolution microscopy
\citep{bartholomew1980changes}. The crystal orientation is precisely
determined and oriented to the required direction, mostly to the \{1,1,1\},
\{1,1,0\} or \{1,0,0\} by the X-ray Laue method. This is followed
by further XRD analysis for both final precise crystallographic orientations,
as well as an initial assessment of the crystalline quality and microstructure.
X-Ray Rocking Curve technique is applied in case of high crystalline
quality assessment. Imaging methods such as optical microscopy, SEM
and STEM, are carried out as well, for determining the physical microstructure
that plays a significant role at the final X-Ray analysis. XRD peak
shift, peak broadening and peak asymmetry are indicative of deformation,
growth fault, twin fault and several additional crystalline disorders
in the FCC metal element crystal (Aluminum single crystal in the current
study). Detailed theoretical calculations and experimental results
are described in \citep{warren1990x}. The FCC structure is a closed
packed structure in which the atoms are arrange in a three layer sequence
resulting in a dense geometric packing. Figure \ref{fig:Representation-of-three}
illustrates a 3D arrangement of three such layers. 
\begin{figure}[H]
\begin{centering}
\begin{tabular}{cc}
\includegraphics[scale=0.27]{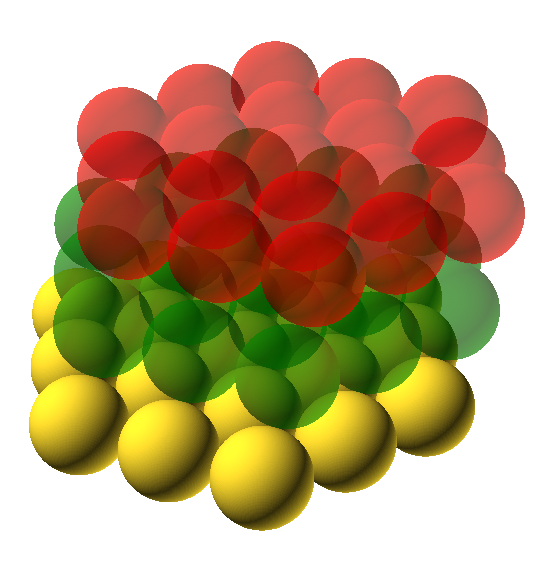} & \includegraphics[scale=0.24]{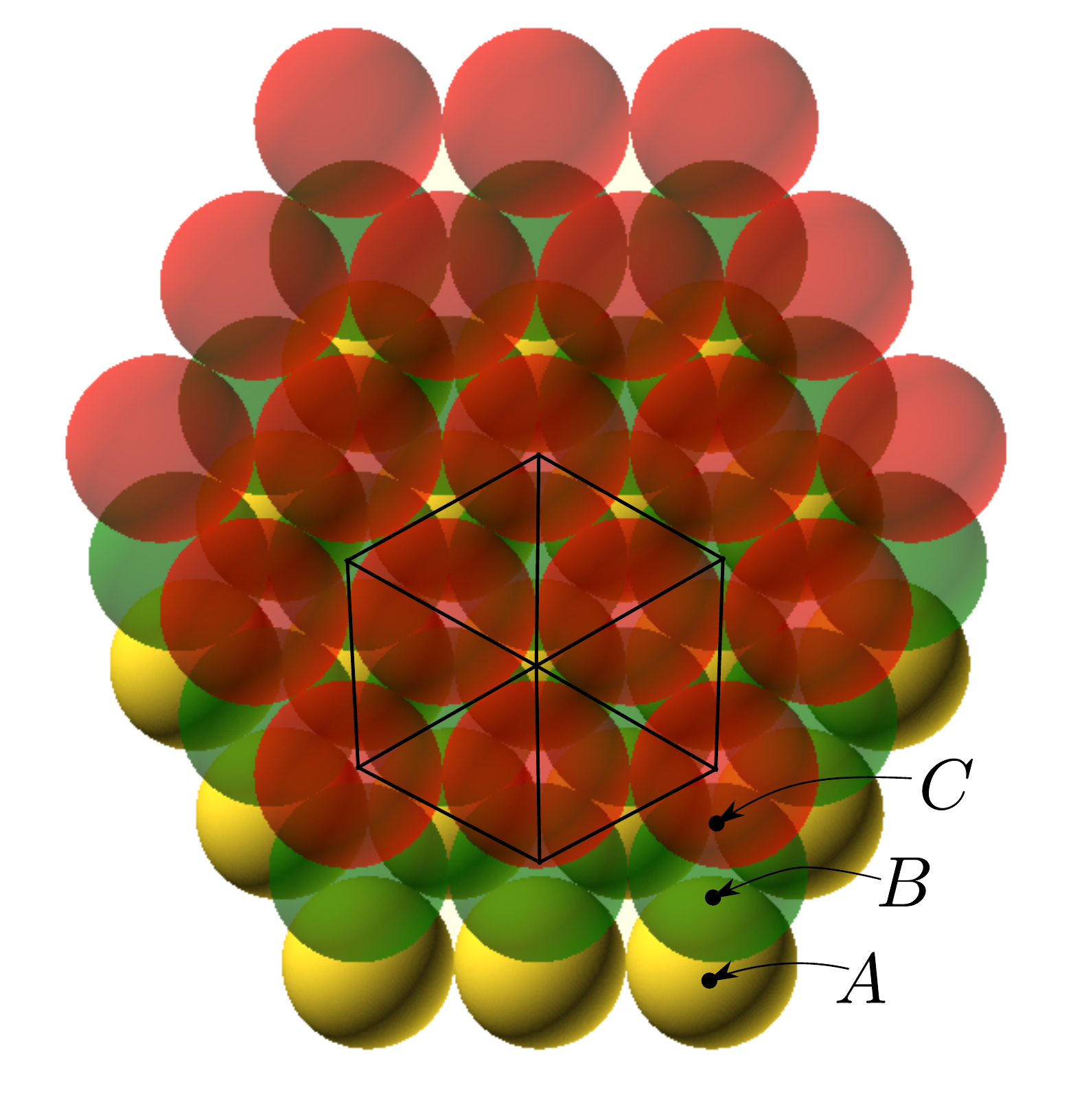}\tabularnewline
(a) & (b)\tabularnewline
\end{tabular}
\par\end{centering}
\centering{}\caption{\label{fig:Representation-of-three}Representation of layers forming
the FCC structure. (a) The recurring layer placement required for
the FCC structure. (b) The actual location of the atoms missing the
apex atom which is part of a top A layer (not shown).}
\end{figure}

\begin{figure}[H]
\centering{}%
\begin{tabular}{cc}
\includegraphics[scale=0.27]{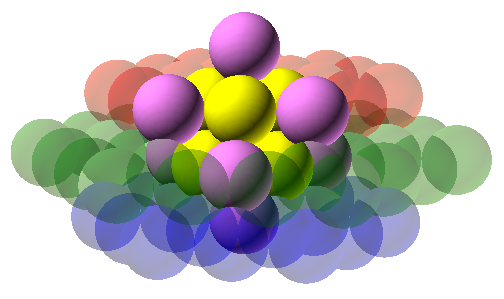} & \includegraphics[scale=0.25]{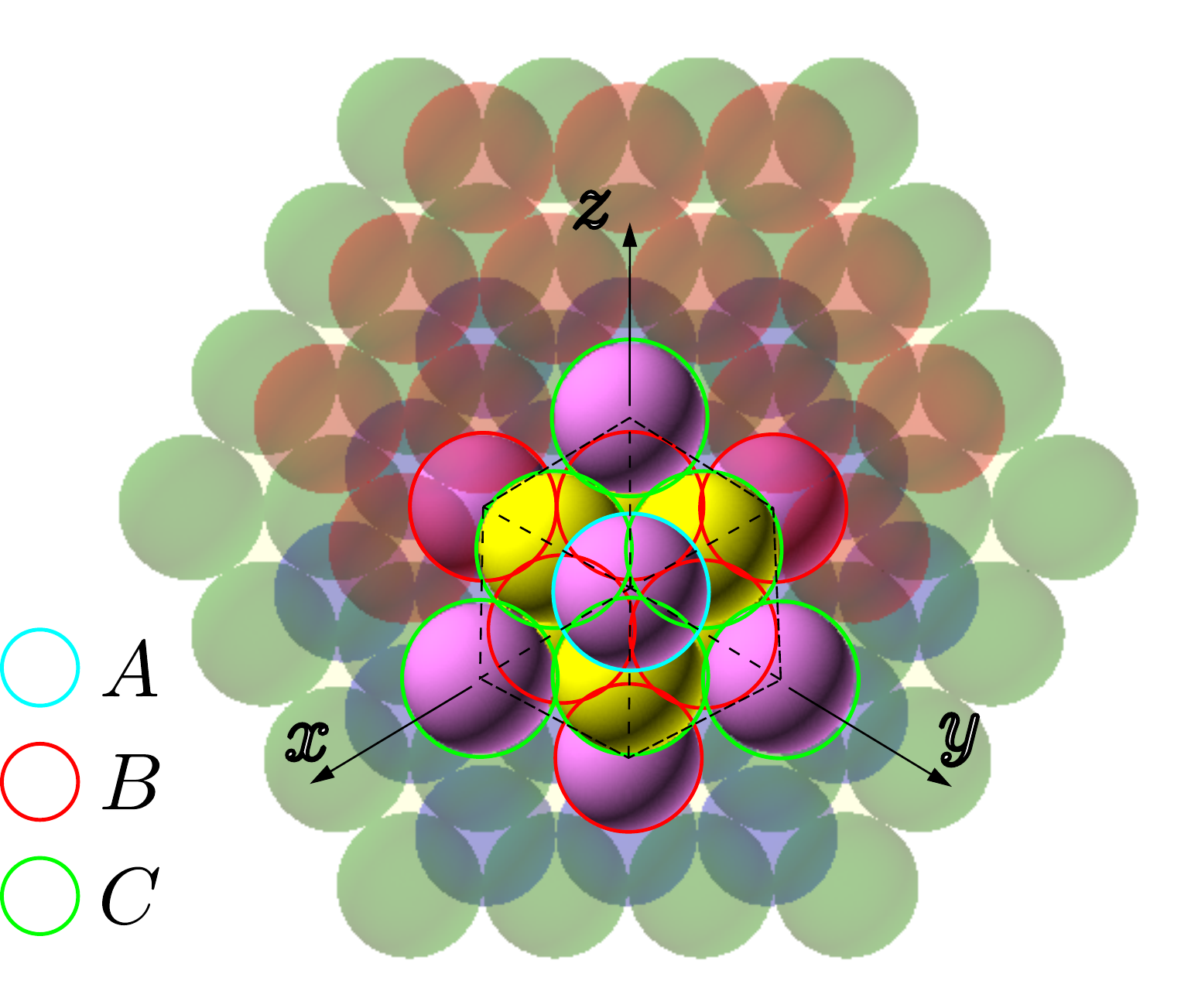}\tabularnewline
(a) & (b)\tabularnewline
\end{tabular}\caption{\label{fig:An-FCC-cell}A FCC unit cell within the layers. (a) Side
view of the FCC cell, (b) View of the cell from the \{1,1,1\} direction. }
\end{figure}

The figure illustrates the layers of the atoms and the relative placement
of each layer providing for the most compact filling. In Fig. \ref{fig:Representation-of-three}b
the orientation of the FCC cell within the 3 layer framework is shown.
As the image is missing a top A layer the apex atom is missing. The
cell structure is a bit obscure and one has to imagine the structure
from the vertices. Figure \ref{fig:An-FCC-cell} elucidates the FCC
cell within the layers.

\begin{figure}[h]
\raggedright{}%
\begin{tabular}{cc}
\includegraphics[scale=0.3]{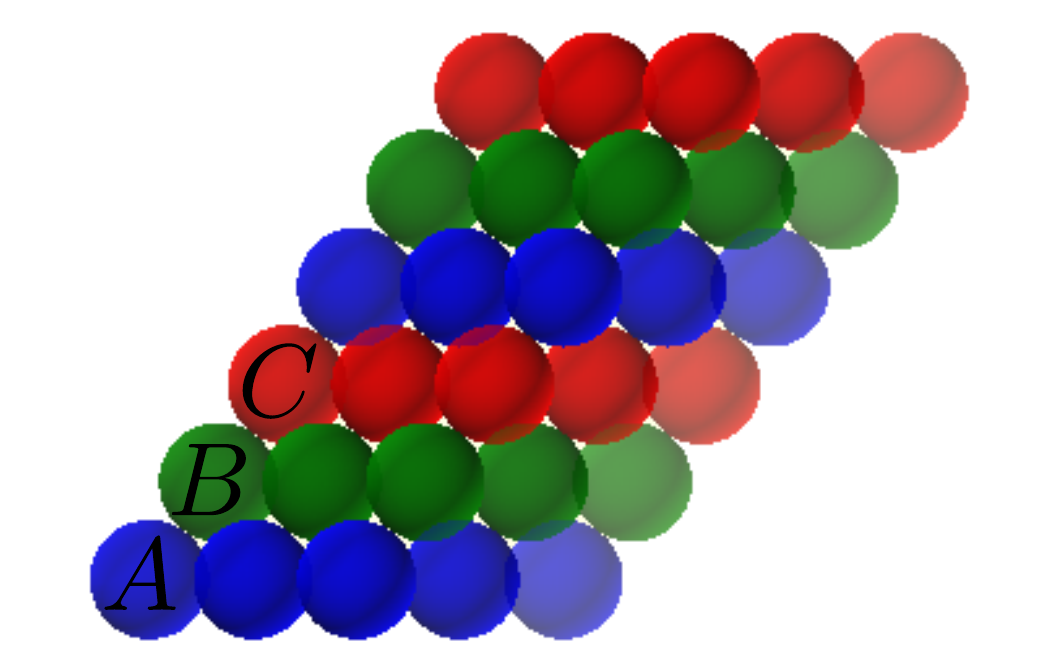} & \includegraphics[scale=0.25]{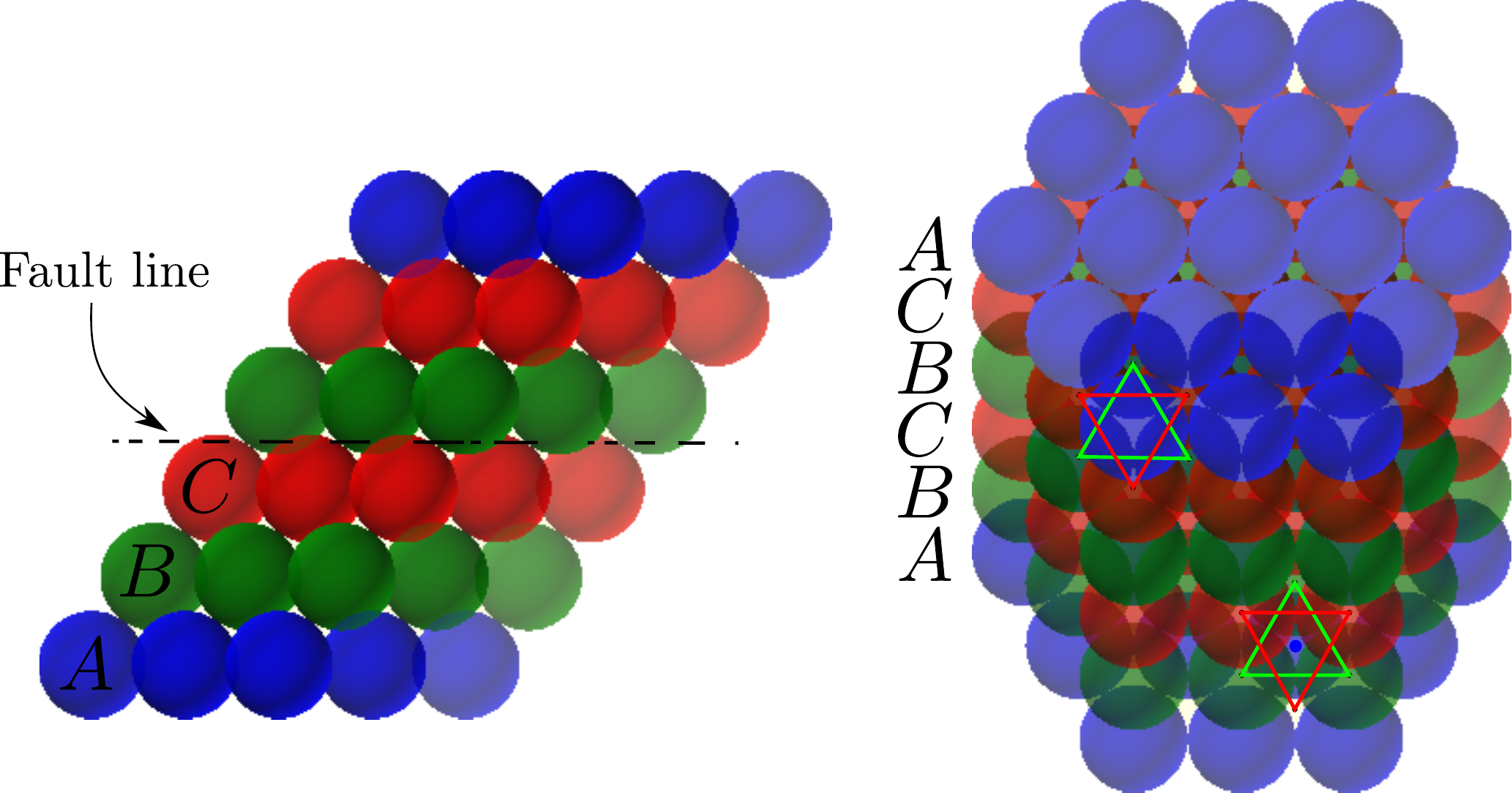}\tabularnewline
(a) & (b)\tabularnewline
\end{tabular}\caption{\label{fig:Layer-stacking-with}Layer stacking with a missing layer.
(a) Regular 3 layer compact stacking resulting in the FCC strucure.
(b) Missing A layer (for example), resulting in the FCC cells missing
the top or bottom atom creating a fault line. }
\end{figure}

\begin{figure}[h]
\centering{}\includegraphics[scale=0.25]{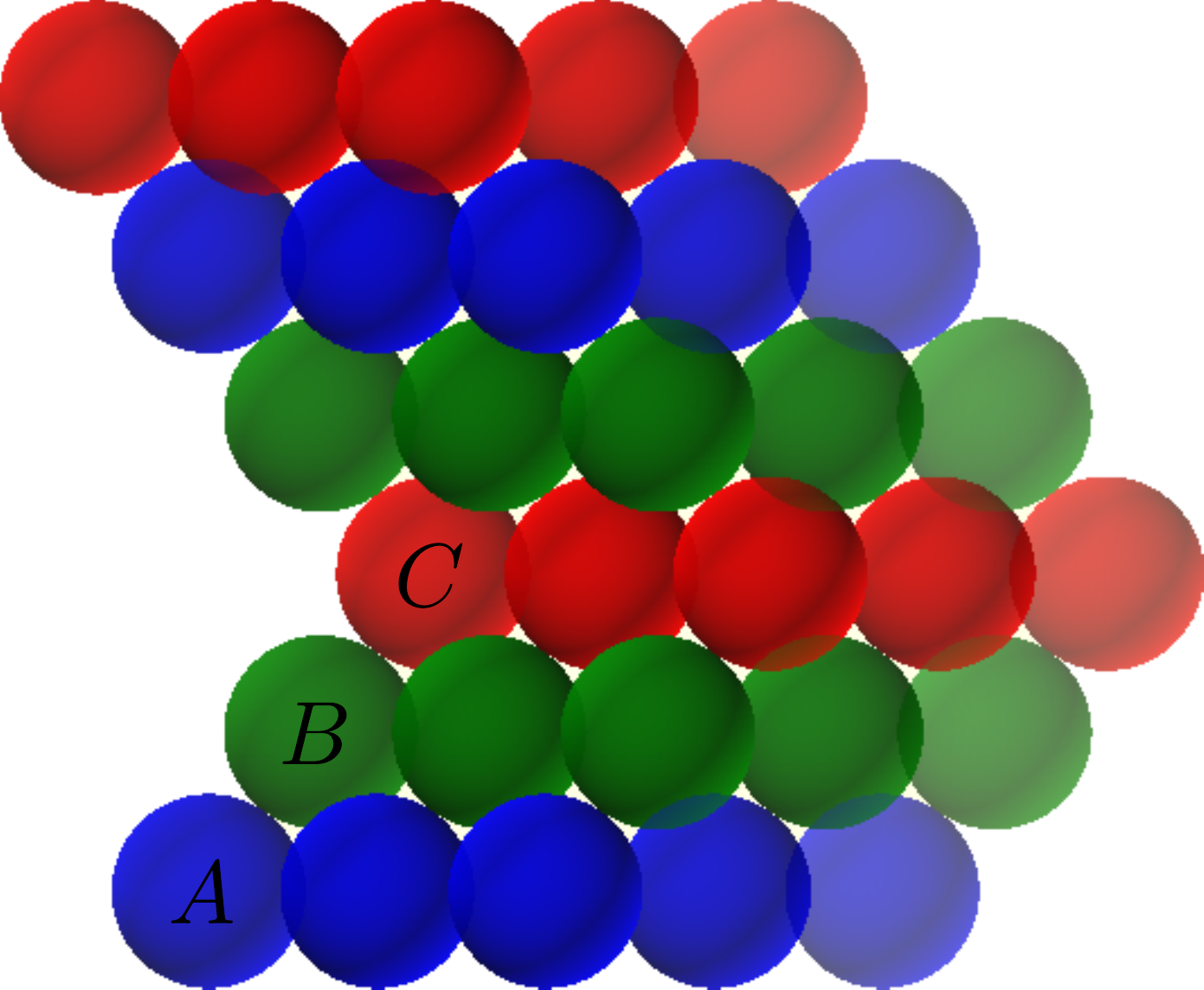}\caption{\label{fig:Layer-stacking-fault-twins}Layer stacking fault where
two adjacent layers switch positions. This results in a bifurcation
where the layers above the fault are a mirror image of the layers
below it.}
\end{figure}

The vertices atoms are illustrated by the violet spheres while the
face centered atoms by the yellow spheres. As may be seen the apex
atom and the bottom atom belong to the A layer placing them one over
the other. Following the perpendicular unit cell axis one can observe
the \{1,1,1\} orientation in Figure \ref{fig:An-FCC-cell}b. The red
bordered atoms highlight the vertex and face atoms at the lower layer,
while the green bordered atoms highlight the ones of the layer above
it. The vertices atoms of each layer form a large triangle bordering
a smaller triangle created by the face atoms of that specific layer,
with the smaller triangle being at a $180\degree$ angle to the one
formed by the vertices. The triangles formed by the atoms in each
layer are a mirror image of the other. Given we had a perfect crystal,
processing it mechanically will introduce defects. Point defects having
less long range influence on the crystals structure. On the other
hand, layer movement along the slip planes due to stress, with the
primary ones being along the \{1,1,1\} direction have much more influence
for e.g. the line defects, the edge and screw dislocations. More subtle,
but of significant influence on the diffraction patterns, are the
stacking faults resulting in fault lines Fig. \ref{fig:Layer-stacking-with}
and twins Fig. \ref{fig:Layer-stacking-fault-twins}

Stacking faults in which one layer is missing, results in an alternating
sequence of two layered stacks which result in a hexagonal closed
packed structure along that layer. It could also be viewed as the
top and bottom FCC cells missing one apex or base atom. Another stacking
fault is when two adjacent layer switch positions as in Fig. \ref{fig:Layer-stacking-fault-twins}

\begin{figure}[h]
\begin{centering}
\includegraphics[scale=0.32]{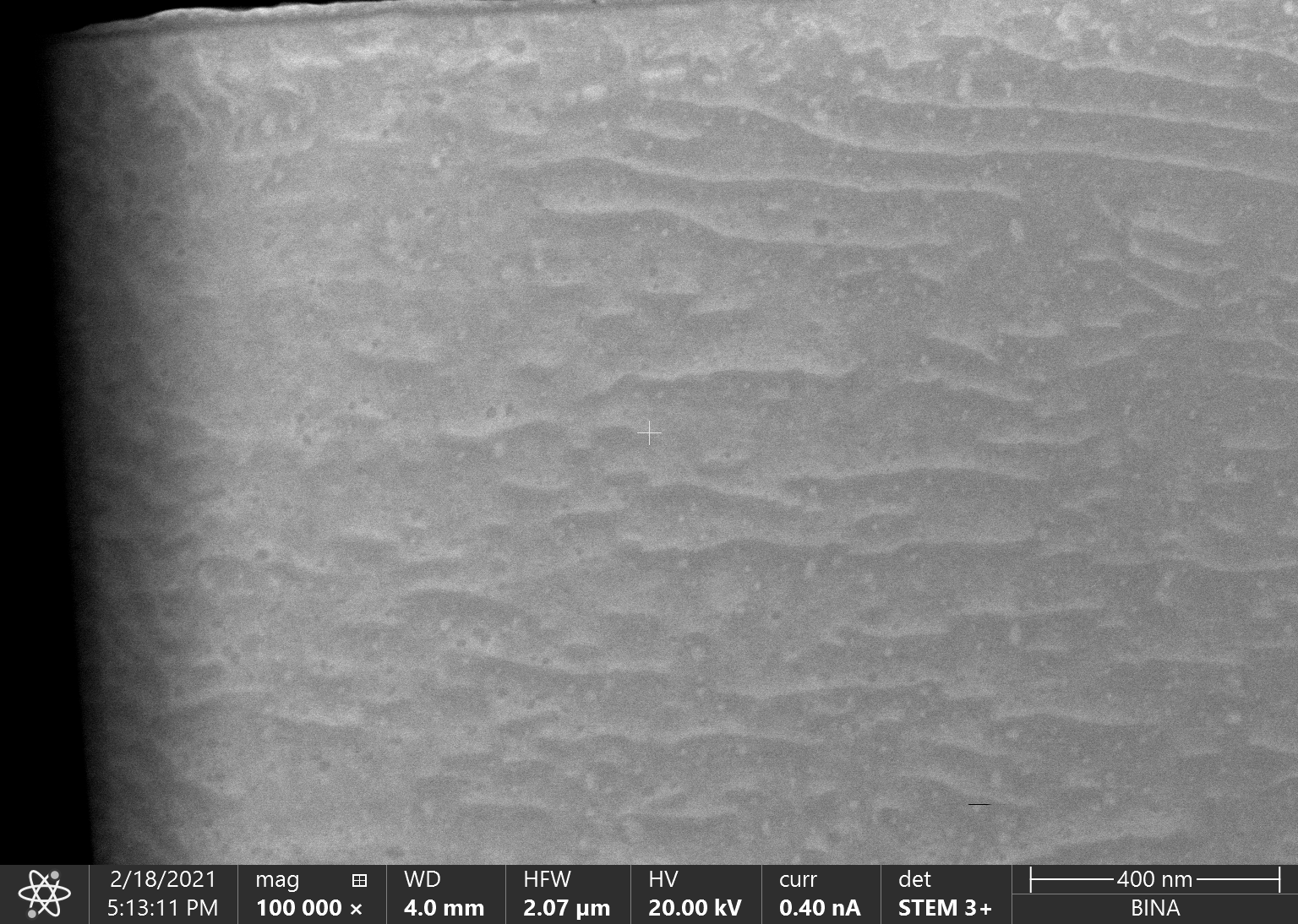}
\par\end{centering}
\caption{\label{fig:STEM-image-of-low-angle-boundary}A $\times80,000$ STEM
image of low angle tilt boundaries, creating what is referred to as
mosaic structures in aluminum.}
\end{figure}

Such an occurrence leads to a bifurcation with the layers on both
sides of the fault proceed as mirror images of one another prompting
the growth of two crystals with a common layer in two different directions.
Such a fault may have a considerable effect on the peak broadening.
As we are dealing with single crystals we do not expect to observe
grain boundaries, but low angle tilt boundaries resulting from dislocations
and faults are present and its correlation with the rocking curve
peak broadening in a manner not yet fully understood. They are observed
using STEM imaging as can be seen in Fig. \ref{fig:STEM-image-of-low-angle-boundary}.

Those boundaries differ from grain boundaries as from their very essence
the bulk has a growth orientation and only locally differ from this
orientation due to a periodic array of dislocations and stacking faults.
The degree of this misorientation may be observed by the rocking curve
method and in future published work will be correlated with the electrical
conductivity in aluminum.

\section{Experimental Results}

\begin{figure}[h]
\centering{}\includegraphics[scale=0.32]{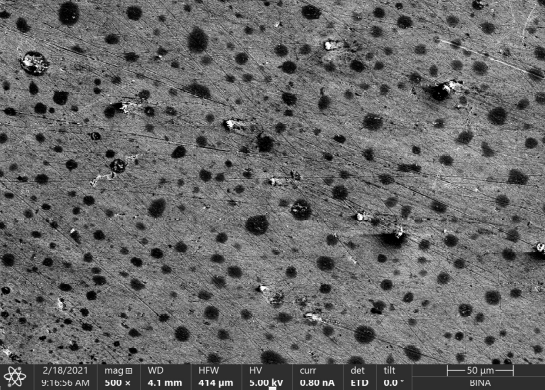}\caption{\label{fig:SEM-image-of}SEM image of the aluminum surface after cutting.
The high pitting density, with the pit sizes reaching 20nm.}
\end{figure}

Following the Bridgman crystal Growth of Aluminum single crystals,
samples were prepared for the purpose of crystallographic orientation,
XRD analysis, X-Ray Rocking Curve analysis and crystalline quality
evaluation by STEM imaging. As the material is very soft and ductile
initial fabrication was conducted using electro-erosion. Figure \ref{fig:SEM-image-of}
illustrates the result of an aluminum sample after it was cut. The
STEM image displays the highly damaged top surface as observed in
figure \ref{fig:SEM-image-of}, which reduces the X-Ray intensity
and increases the peak broadening. 

ICP-AES measurements indicated that trace elements of brass exist
in the samples. Table \ref{tab:ICP-AES-results-of} illustrates the
trace elements within a sample compared to the trace elements found
within the starting material and reference. 
\begin{table}[h]
\begin{centering}
\resizebox{9cm}{!}{%
\begin{tabular}{l>{\raggedleft}p{0.4cm}@{\extracolsep{0pt}.}p{0.4cm}>{\raggedleft}p{0.4cm}@{\extracolsep{0pt}.}p{0.4cm}>{\raggedleft}p{0.4cm}@{\extracolsep{0pt}.}p{0.4cm}>{\raggedleft}p{0.4cm}@{\extracolsep{0pt}.}p{0.4cm}>{\raggedleft}p{0.4cm}@{\extracolsep{0pt}.}p{0.4cm}>{\raggedleft}p{0.4cm}@{\extracolsep{0pt}.}p{0.4cm}>{\raggedleft}p{0.4cm}@{\extracolsep{0pt}.}p{0.4cm}>{\raggedleft}p{0.4cm}@{\extracolsep{0pt}.}p{0.4cm}}
\hline 
 & \multicolumn{2}{p{0.8cm}}{\textbf{Ag}} & \multicolumn{2}{p{0.8cm}}{\textbf{Ca}} & \multicolumn{2}{p{0.8cm}}{\textbf{Cu}} & \multicolumn{2}{p{0.8cm}}{\textbf{Fe}} & \multicolumn{2}{p{0.8cm}}{\textbf{Mg}} & \multicolumn{2}{p{0.8cm}}{\textbf{Zn}} & \multicolumn{2}{p{0.8cm}}{\textbf{Cr}} & \multicolumn{2}{p{0.8cm}}{\textbf{S}}\tabularnewline
\hline 
\textbf{Starting material} & 3&914 & \multicolumn{2}{p{0.8cm}}{-} & \multicolumn{2}{p{0.8cm}}{-} & 6&06 & \multicolumn{2}{p{0.8cm}}{-} & \multicolumn{2}{p{0.8cm}}{-} & \multicolumn{2}{p{0.8cm}}{-} & 128&2\tabularnewline
\textbf{Reference} & 3&98 & \multicolumn{2}{p{0.8cm}}{-} & \multicolumn{2}{p{0.8cm}}{-} & 6&45 & \multicolumn{2}{p{0.8cm}}{-} & \multicolumn{2}{p{0.8cm}}{-} & 0&12 & 87&59\tabularnewline
\textbf{Al EDM cut} & \multicolumn{2}{p{0.8cm}}{-} & 132&49 & 495&06 & 141&41 & 13&02 & 170&64 & 0&66 & \multicolumn{2}{p{0.8cm}}{-}\tabularnewline
\hline 
\end{tabular}}
\par\end{centering}
\caption{\label{tab:ICP-AES-results-of}ICP-AES results of several samples
(PPM)}
\end{table}
 Post EDM cutting displays obvious incorporation of brass elements
into the crystal material. Following this SEM imaging, a STEM sample
was prepared using a FIB process with the cut perpendicular to the
observed surface in Fig. \ref{fig:SEM-image-of}.

\begin{figure}[h]
\begin{centering}
\includegraphics[scale=0.32]{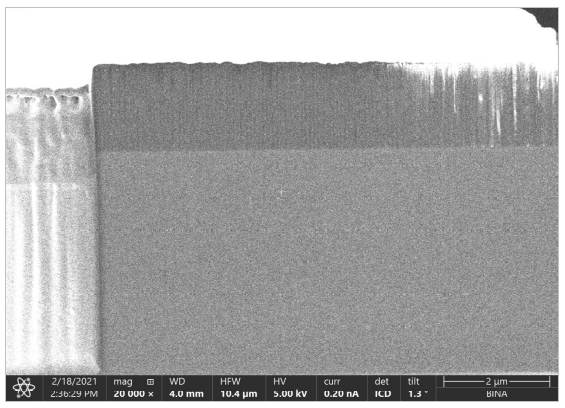}
\par\end{centering}
\caption{A low resolution STEM cross section image of the sample close to its
surface. The damage to the machined layer can be clearly see, having
the approximate thickness of $1.8\mu m$.}
\end{figure}

\begin{figure}[h]
\begin{centering}
\includegraphics[scale=0.32]{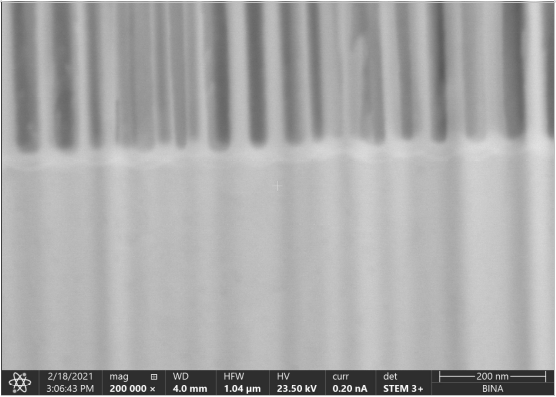}
\par\end{centering}
\caption{\label{fig:High-resolution-STEM}High resolution STEM image of the
lamella prepared using the FIB technique}
\end{figure}

A magnified higher resolution STEM image (Fig. \ref{fig:High-resolution-STEM}),
displays column structure (``membrane like'') cavities with diameters
corresponding to the previous pits observed in figure \ref{fig:SEM-image-of}. 

The image illustrates how the surface damage extends deep into the
material altering its structure. This damage was found to propagate
to a depth of approximately 2 mm with similar results reported in
the past by Gorman et. al. \citep{gorman1969mobility}. Such damage
results in reduced XRD intensity and considerable peak broadening.
Those defects within the aluminum and on its external surfaces scatter
conducting electrons and contribute to the resistivity. As the dimensions
of the conductors are reduced, which is the case in the solar cell,
the electrons interacting with the defects increase considerably resulting
in a rise in resistivity. Figure \ref{Fig1_XRD_non_single} below,
illustrates an example of the XRD data that was acquired from a crystal
with relatively poor quality. As may be observed, the diffraction
pattern shows low peak intensity and mixed crystallographic orientation
data. 

\begin{figure}[h]
\centering{}\includegraphics[scale=0.45]{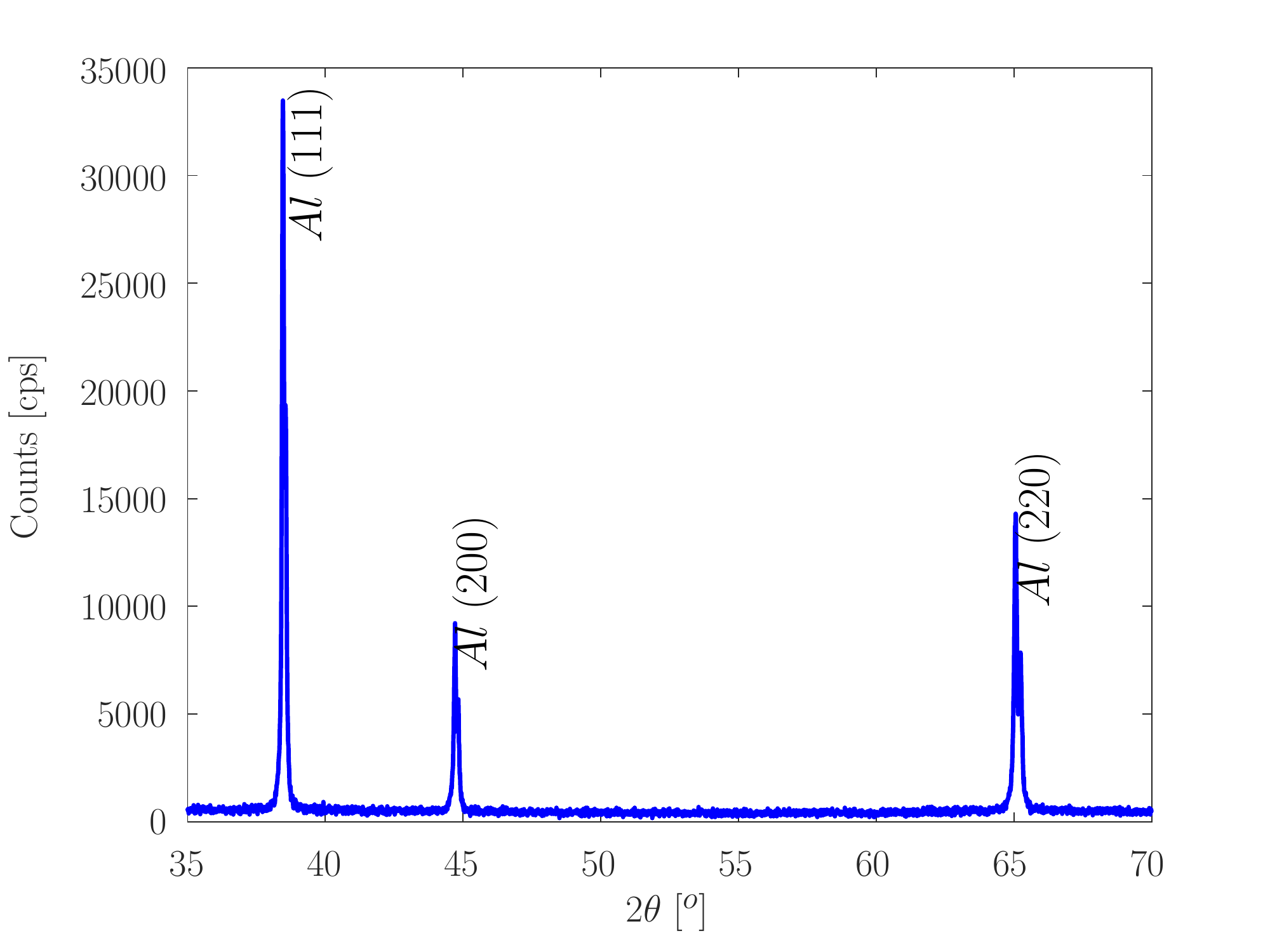}\caption{\label{Fig1_XRD_non_single}An example of \{1,1,1) oriented Aluminum
Crystal. Minor diffraction at the \{2,0,0\} and \{2,2,0\} orientations
are also visible. }
\end{figure}

Figure \ref{Fig2_XRD_single} below, shows an example of the XRD data
that was acquired from a crystal of high quality and lattice perfection.
A very high peak intensity at the range of 9e7 (20 millions counts),
with a single distinguished peak (disregarding $K_{\beta}$) diffraction
in the \{1,1,1\} crystallographic orientation. 

\begin{figure}[h]
\centering{}\includegraphics[scale=0.45]{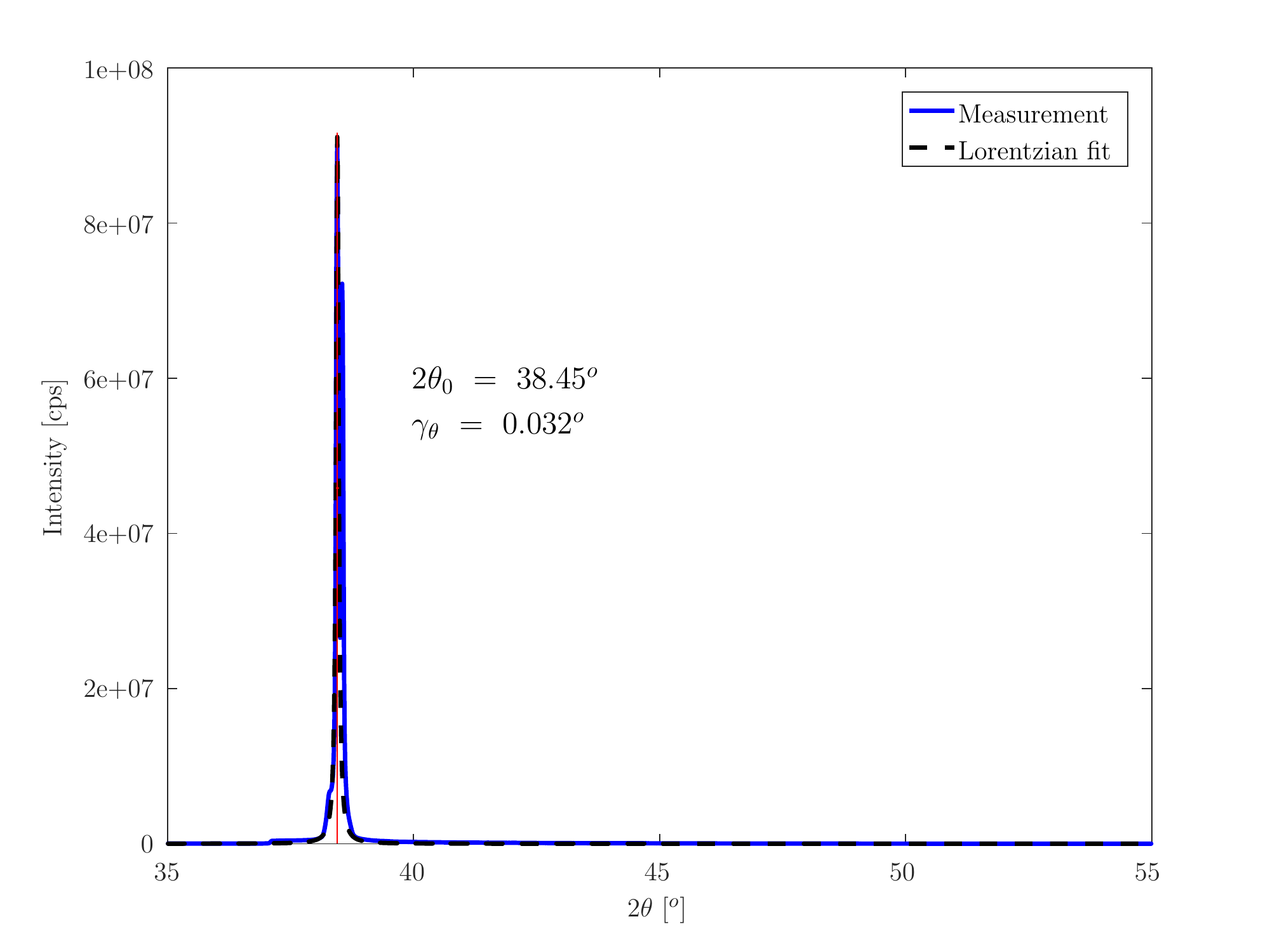}\caption{\label{Fig2_XRD_single}An example of an XRD chart for a precisely
oriented Aluminum Crystals in the \{1, 1, 1\} direction. }
\end{figure}

\begin{figure}[h]
\centering{}\includegraphics[scale=0.45]{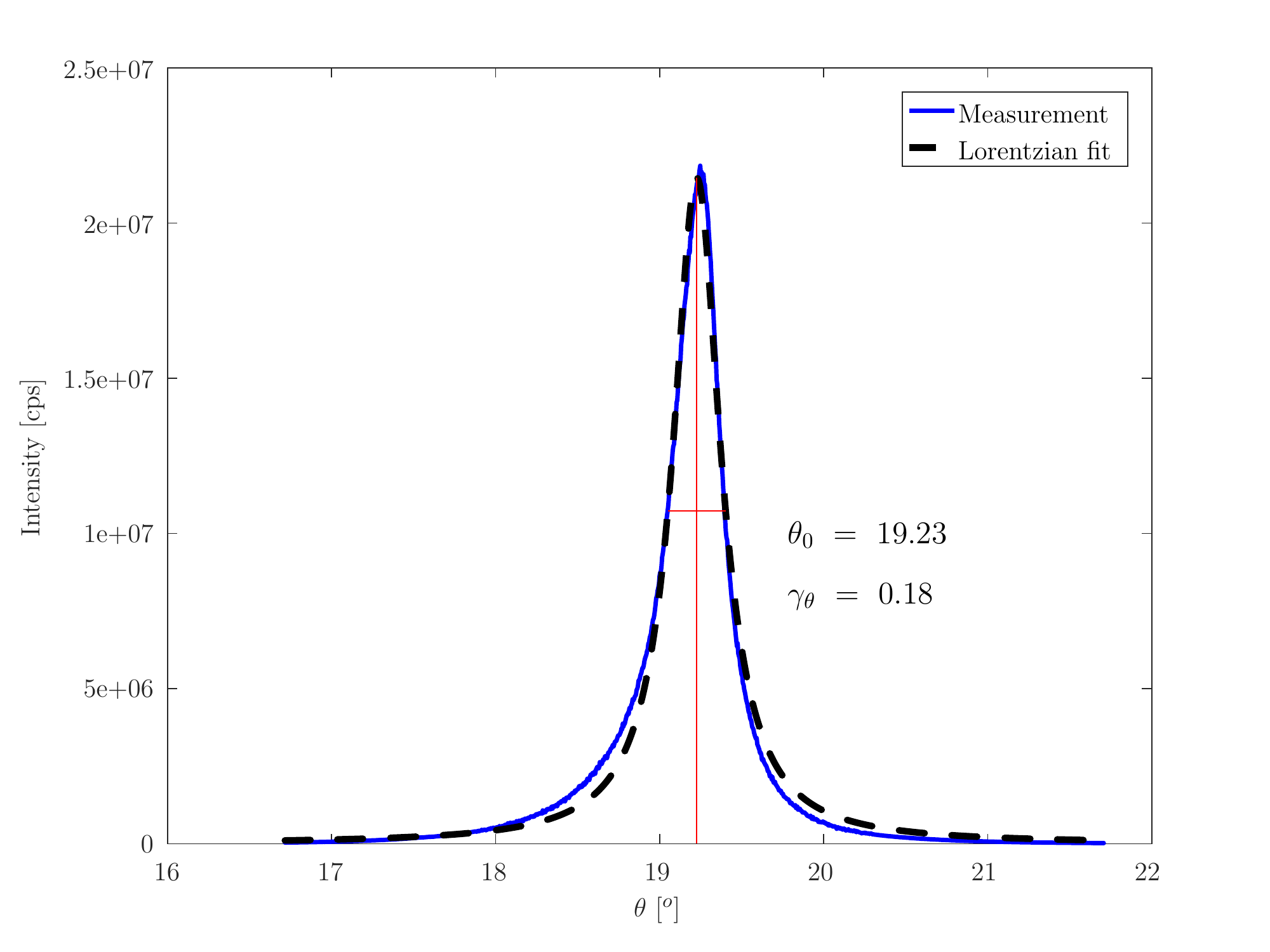}\caption{\label{Fig3_XRD_Rocking_curve}An example of an X-Ray Rocking Curve
chart, with a FWHM of about 0.36 degrees and very high peak intensity
at the range of 2e7, a slight asymmetry is evident as well. }
\end{figure}

Next, figure \ref{Fig3_XRD_Rocking_curve} shows the Rocking Curve
peak, with a FWHM of about 0.36 degrees, yet, it should be noted that
the RC shape exhibits a minor asymmetry, indicating on some crystalline
in-homogeneity. As can be seen in the figure, there is an axial divergence
at the lower angles of the peak, causing asymmetry. This asymmetry
is an indication of an anisotropic strain.

\begin{figure}[H]
\begin{centering}
\includegraphics[scale=0.72]{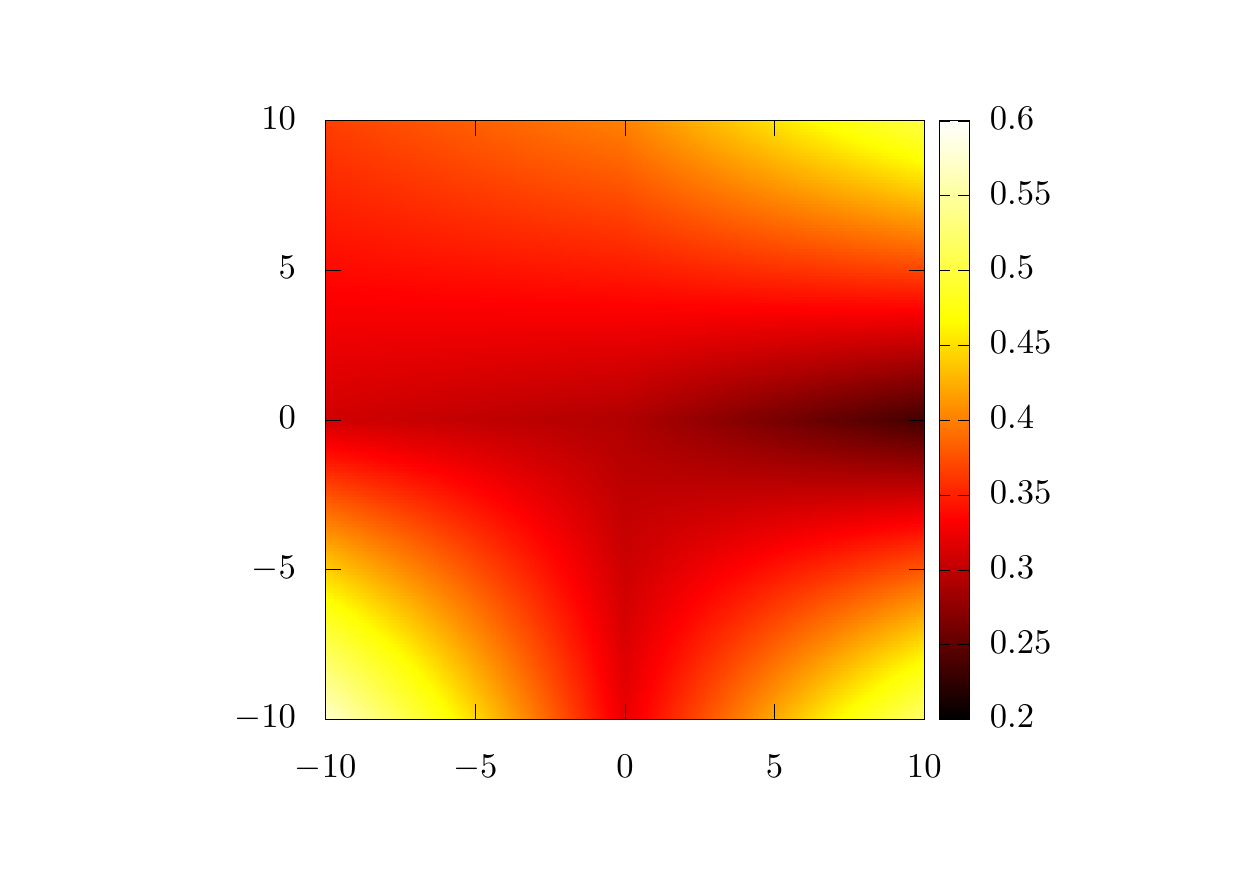}
\par\end{centering}
\caption{\label{fig:Rocking-curve-mapping}Rocking curve mapping of the tilt
spread on a 2x2 cm area from a crystal cross section.}

\end{figure}

\begin{figure}[H]
\begin{centering}
\includegraphics[scale=0.72]{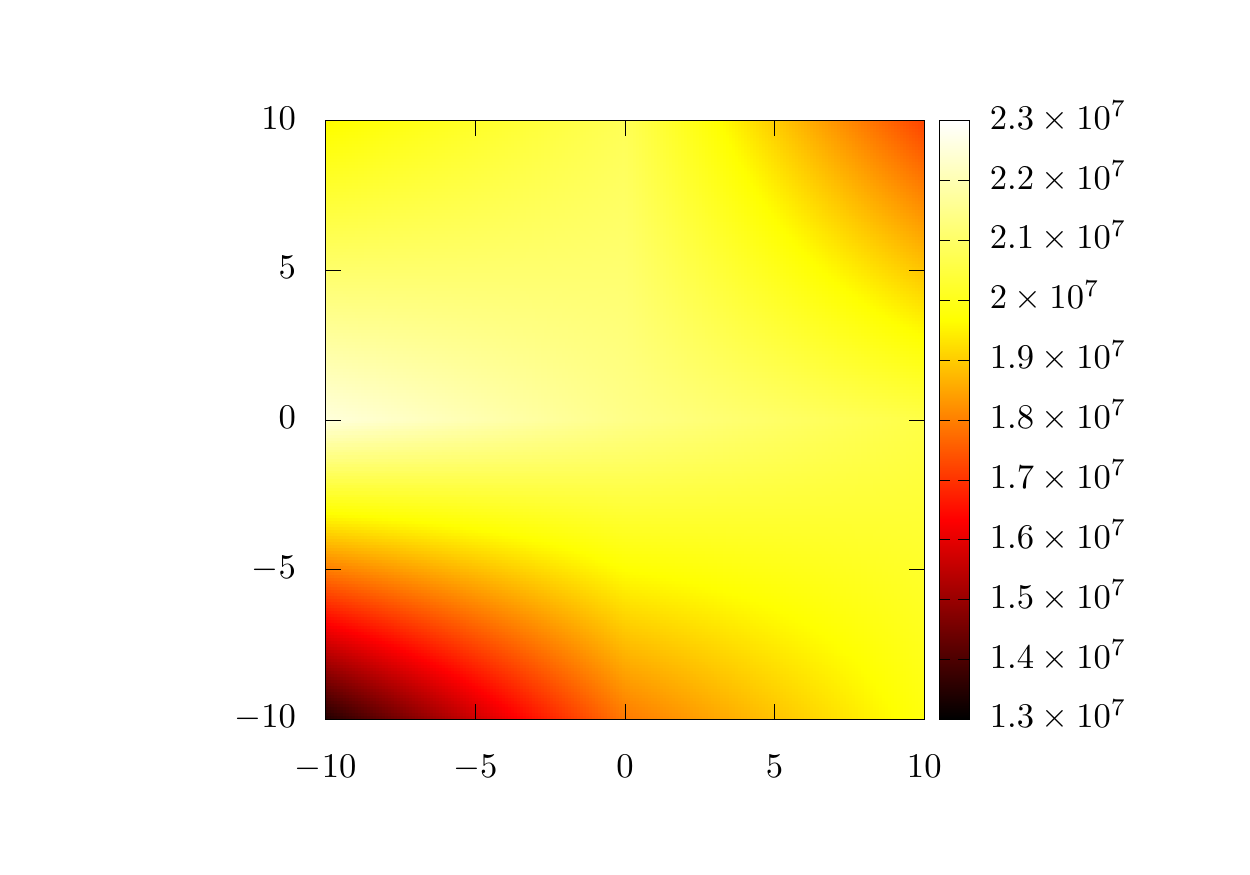}
\par\end{centering}
\caption{\label{fig:Rocking-curve-mapping-peak-intensity}Rocking curve mapping
of peak intensity on a 2x2 cm area from a crystal cross section}

\end{figure}

\begin{figure}[h]
\begin{centering}
\includegraphics[scale=0.5]{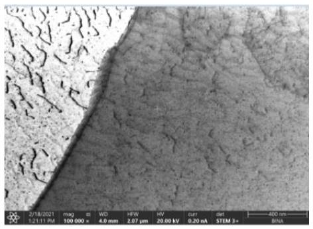}
\par\end{centering}
\begin{centering}
(a)
\par\end{centering}
\begin{centering}
\includegraphics[scale=0.52]{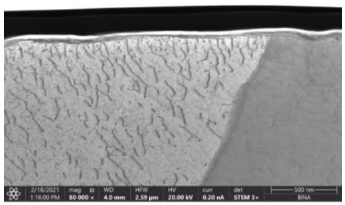}
\par\end{centering}
\centering{}(b)\caption{\label{Fig4_STEM_images_1}(a) is a STEM image of an Aluminum crystal
with high defect density, reflected at the observed sub grains and
dislocation defects. (b) Shows displays the native oxide layer of
the surface of the crystal. }
\end{figure}

The results were fit using a Lorentz (Cauchy) PDF (probability distribution
function), that was found to better fit the results. If we take the
Lorentzian's FWHM as a measure of orientation spread of the mosaic
structure tilt, we can estimate that it is approximately $0.18^{o}$.
Following the FWHM angle spread of the tilt angles we can calculate
the upper limit of the dislocation density \citep{gay1953estimation}.
\begin{equation}
D=\frac{(FWHM/2)^{2}}{9b^{2}}=\frac{\gamma_{\theta}^{2}}{9b^{2}}\label{eq:dislocation_density}
\end{equation}
where $\gamma_{0}$ is the tilt spread, and b is the Burgers vector.
For the aluminum with an FCC structure in the (1,1,1) direction, the
Burgers vector is given by
\begin{equation}
b=\frac{a}{2}\cdot\sqrt{(h^{2}+k^{2}+l^{2})}=\frac{a\sqrt{3}}{2}\label{eq:burgers vector}
\end{equation}
where a is the side length of the unit cell. As it is $4.046\times10^{-10}\,m$,
$b=3.504\times10^{-10}\,m$. Thus the upper limit estimate of the
dislocation density is 
\[
D=8.93\times10^{8}\,[cm^{-2}]
\]
which is considered high compared to work published by Mizuno Et al.
\citep{mizuno2005vacancy}. This is reasonable as in the mentioned
work, ultra high purity raw materials were used (7N), the samples
were thin with an elaborate annealing process. Considering the stress
within the crystal, we could expect that the stress is non uniform.
This was evident by rocking curve mapping resulting in different tilt
spreads as can be seen in Figure \ref{fig:Rocking-curve-mapping}.
as the tilt spread is correlated to the dislocation density which
is further correlated to the stress, it shows the non uniform stress
within the sample. This must reflect on the peak intensity (counts)
as shown in Figure \ref{fig:Rocking-curve-mapping-peak-intensity}.
While the XRD rocking curve angle spread is an average of many low
tilt angle subgrains, being macroscopic in nature, the fine structure
can be observed using STEM imaging. Figures \ref{Fig4_STEM_images_1}a
and \ref{Fig4_STEM_images_1}b, illustrate the STEM results displaying
the dislocation defects in a crystal that exhibits sub grain boundaries
(low angle) within grains, and tilt boundaries that may significantly
differ from each other. 

\section{Conclusions}

X-ray diffraction methods and Rocking-curve analysis, supported by
SEM and STEM imaging, were shown to display, in an effective manner
the three dimensional crystalline quality of aluminum single crystals.
This is demonstrated with both poor and high quality aluminum single
crystals grown using the veteran Bridgman method. Growth and fabrication
of the crystals were still found to be short of the strived state
of the art. It was further demonstrated that sample fabrication by
rough electro-erosion cutting, exhibited a membrane like, porous deformed
structure at the aluminum as cut surface. Supplementing electrical
conductivity measurements of aluminum, quality assessment of defects
in front cell aluminum conductors can assist in designing novel low
resistance aluminum conductors replacing the currently widely used
and relatively rare silver.

\bibliographystyle{unsrt}
\bibliography{references}

\end{document}